\documentclass[review]{elsarticle}

\usepackage{lineno}

\usepackage{multirow,setspace,amssymb,amsmath,graphicx,color,rotating,subfigure,url}
\usepackage{lineno}
\usepackage{natbib}
\usepackage{textcomp}
\usepackage{CJK}
\usepackage{bm}%
\usepackage{rotating}
\usepackage{diagbox}
\usepackage{epsfig}
\usepackage{dcolumn}
\usepackage{epstopdf}
\usepackage{natbib}
\biboptions{sort&compress}
\usepackage[hidelinks]{hyperref}

\begin{document}

\begin{frontmatter}

\title{Vital node identification in hypergraphs via gravity model}

\author[inst1]{Xiao-Wen Xie}
\author[inst1]{Xiu-Xiu Zhan\corref{ca1}}\ead{zhanxiuxiu@hznu.edu.cn}
\author[inst2]{Zi-Ke Zhang}
\author[inst1]{Chuang Liu\corref{ca1}}\ead{liuchuang@hznu.edu.cn}
\address[inst1]{Alibaba Research Center for Complexity Sciences, Hangzhou Normal University, Hangzhou, 311121, P. R. China}
\address[inst2]{College of Media and International Culture, Zhejiang University, Hangzhou 310058, PR China}
\cortext[ca1]{Corresponding authors.}

\begin{abstract}
Hypergraphs that can depict interactions beyond pairwise edges have emerged as an appropriate representation for modeling polyadic relations in complex systems. With the recent surge of interest on researching hypergraphs, the centrality problem has attracted abundant  attention due to the challenge of how to utilize higher-order structure for the definition of centrality metrics. 
In this paper, we propose a new centrality method (HGC) on the basis of gravity model as well as a semi-local HGC (LHGC) which can achieve a balance between accuracy and computational complexity.
Meanwhile, two comprehensive evaluation metrics, i.e.,  a complex contagion model in hypergraphs which mimics the group influence during the spreading process and network $s$-efficiency based on the higher-order distance between nodes, are first proposed to evaluate the effectiveness of our methods.
The results show that our methods can filter out nodes that have fast spreading ability and are vital in terms of hypergraph connectivity. 

\end{abstract}


\end{frontmatter}

\section{Introduction}
\label{Intro}
\makeatletter
\newcommand{\rmnum}[1]{\romannumeral #1}
\newcommand{\Rmnum}[1]{\expandafter\@slowromancap\romannumeral #1@}
\makeatother

Targeting vital nodes in the networks, which is also referred to as centrality problem, aims to assign the nodes scores that quantifies their importance, so that one can identify important nodes and optimize the allocation of resources.
During the past decades, researches on centrality have garnered immense applications in various domains, including disease transmission~\cite{bell1999centrality, zeng2021identifying, liu2020computational}, political propagation~\cite{aghdam2016opinion}, rumor suppression~\cite{shah2011rumors, ilyas2011identifying}, advertising~\cite{mochalova2014targeted} and traffic governance~\cite{du2014new}. 
These studies mostly focused on the pairwise interactions to characterize the relationship between individuals~\cite{albert2002statistical, cimini2019statistical, boccaletti2006complex}.
However, we shall not neglect that higher-order interactions that enclose multiple individuals are more general in the real systems, such as chatting groups, protein complexes and so on~\cite{ramadan2004hypergraph, lung2018hypergraph}.
Simple as the ordinary network is, it cannot depict group effects among multiple nodes. 
In contrast, hypergraph, which allows a hyperedge to connect multiple nodes~\cite{mayfield2017higher}, can nicely compensate for the shortcomings of the ordinary network representation of a complex system.

The centrality in hypergraphs can be divided into two categories, namely the centrality of a node or a hyperedge~\cite{kapoor2013weighted, tudisco2021node}.
Degree centrality holds a simple idea that the node with a larger number of neighbors is more influential. 
Since it assumes that two nodes are mutually adjacent if they exist in the same hyperedge, it is an equivalence of degree in an ordinary network.
Compared to degree, hyperdegree is defined as the number of hyperedges that a node belongs to, and takes the higher-order information into consideration~\cite{berge1973graphs}.
Both methods only measure local influence of a node, while some researchers seek to measure nodes from a broader view by considering global topological characteristics, such as paths and eigenvectors.
Estrada et al.~\cite{estrada2006subgraph} extended the subgraph centrality to hypergraphs, which takes the number of closed loops through a node as its subgraph centrality.
Benson et al.~\cite{benson2019three} proposed an eigenvector centrality that is applicable to uniform hypergraphs whose hyperedges are in uniform size.
Kovalenko et al.~\cite{kovalenko2021vector} defined a vector centrality that describes the importance of nodes in different sizes of hyperedges.
Aksoy et al.~\cite{aksoy2020hypernetwork} defined the $s$-closeness centrality and $s$-eccentricity of a hyperedge in a hypergraph.

In the majority of existing works, the centrality measures in hypergraphs either lack of the characterization of higher-order structures, or are too strict to implement in general.
More importantly, most existing work hasn't considered the higher-order interactions when evaluating the performance of centrality metrics, leaving the effectiveness of the existing methods in hypergraphs unknown.
In this paper, we first define the higher-order distance in a hypergraph to capture the higher-order structures, and further define a centrality measure in hypergraphs based on the gravity model, which is denoted as HGC in the following context.
The method takes into account both local and global structural information of the nodes.
Furthermore, a semi-local centrality measure based on the gravity model (denoted as LHGC) is proposed to reduce the computational complexity. We propose two frameworks to evaluate the effectiveness of our methods, i.e., spreading dynamics and hypergraph connectivity.
Experimental results conducted on empirical hypergraphs generated by real-world data show that the nodes screened by HGC and LHGC have stronger and faster dissemination ability compared with the state-of-the-art baselines. And also, these nodes play an important role in sustaining the higher-order connectivity.

The remainder is organized as follows:
we give the definition of a hypergraph as well as the descriptions of the datasets in Section 2; in Section 3, we introduce the definition of our centrality methods; in Section 4, we explain the benchmark and evaluation metrics;
in Section 5, we evaluate and analyze our methods from different perspectives in conjunction with the baselines; the paper is concluded in Section 6.

\section{Hypergraph and Datasets}
\label{sec:Hypergraph and Datasets}

\subsection{Definition of a Hypergraph}
\label{sec:Definition of a Hypergraph}
We denote a hypergraph with $N$ nodes and $M$ hyperedges as $H=(V,\ E)$, where $V=\{v_1,\ v_2,\cdots,v_N\}$ and $E=\{e_1,\ e_2,\cdots,e_M\}$ are the sets of nodes and hyperedges, respectively. 
A hyperedge $e_m \ (m=1,\cdots,M)$ is a collection of nodes, i.e., $e_m \subseteq V$, indicating the interactions between multiple nodes. We use $k_i$ and $k^H_i$ to represent the degree and hyperdegree of node $v_i$, which are defined as the number of neighbors of $v_i$ and the number of hyperedges that contain $v_i$, respectively. The cardinality of a hyperedge $e_m$ is given by $k^E_m=|e_m|$, indicating the number of nodes in $e_m$.
We can further construct the incidence matrix $B_{N \times M}$ of a hypergraph based on the relationship between nodes and hyperedges. 
Specifically speaking, $B_{i m}=1$ if node $v_i$ belongs to hyperedge $e_m$, otherwise $B_{i m}=0$.
We denote the adjacency matrix of a hypergraph as $A_{N \times N}$.
$A_{ij}$ is equal to $1$ if node $v_i$ and $v_j$ exit in the same hyperedge(s), otherwise it is set to 0. It should be noted that $A_{i i}$ is set to $0$.

\begin{figure*}[htp]
    \centering
    \includegraphics[width=\textwidth]{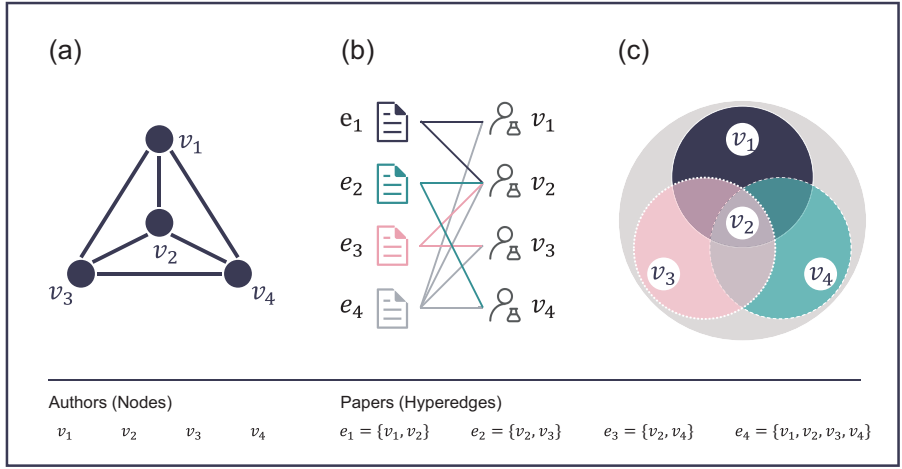}
    \caption{
   Representation of co-authorship in scientific paper collaboration. (a) Ordinary network representation; (b) bipartite network representation; (c) hypergraph representation.
    }
\label{fig:hypergraph}
\end{figure*}

Generally, the hypergraph is superior to other network representations in representing the interactions among more than two entities in the real-world complex systems (see the example illustrated in figure \ref{fig:hypergraph}). 
Taking the co-authorship of scientific papers as an example, we can use three different network representations. 
In figure \ref{fig:hypergraph}(a), we use an ordinary network to represent the co-authorship, in which there is an edge between two authors if they have collaborated at least once. 
figure \ref{fig:hypergraph}(b) is a bipartite network representation, where the left and right columns represent papers and researchers, respectively. 
If a researcher is one of the authors of a paper, there is an edge between the researcher and the paper.
figure \ref{fig:hypergraph}(c) leverages a hypergraph to represent the co-authorship. 
The figures show that researcher $v_1,\ v_3$ and $v_4$ have separately collaborated with $v_2$ in different papers, which can be clearly observed in figure \ref{fig:hypergraph}(b) and (c). 
However, the collaboration in different papers cannot be distinguished in figure \ref{fig:hypergraph}(a). 
That is to say, ordinary network cannot capture how the researchers are collaborated in different papers. In addition, the representation of a bipartite network fails to illustrate the higher-order relationship between the researchers both explicitly and naturally~\cite{battiston2020networks}.

\subsection{Description of the Datasets} 
In the following, we illustrate hypergraphs generated by real-world data from different domains, which will be used to validate the effectiveness of our centrality measures in the subsequent sections. 
We collect $8$ datasets, and the details of each dataset and how hypergraphs are constructed are given as follows:

\renewcommand{\labelitemi}{$\bullet$}
\begin{itemize}
\item \textbf{email-Enron:} A node corresponds to an Enron employee, and each hyperedge consists of the sender and all recipients of an email~\cite{lotito2022higher}.
\item \textbf{Algebra \& Geometry:} A node represents a user, and a hyperedge represents a set of users whose published answers on \url{MathOverflow.net} are tagged with algebra and geometry, respectively~\cite{amburg2020hypergraph}.
\item \textbf{Bars-Rev \& Restaurants-Rev:} A node represents a user of \url{yelp.com}. 
A hyperedge represents a set of users who have posted reviews on bars and restaurants with the same sub-tag, respectively~\cite{amburg2020hypergraph}.
\item \textbf{Music-Rev:} A node represents a user of Amazon, and a hyperedge denotes a set of users who have reviewed products with the same sub-tag of the regional blues tag~\cite{ni2019justifying}.
\item \textbf{NDC-classes:} A node represent a label of classes and a hyperedge represents a drug consists of a set of class labels~\cite{yoon2020much}.
\item \textbf{iAF1260b:} A node represents a metabolite and a hyperedge represents a set of metabolites that are involved in a metabolic reaction~\cite{feist2010model}.
\end{itemize}
We show the topological properties of the hypergraphs generated by the above datasets in Table~\ref{tab:datasets}, where the number of nodes varies from hundreds to thousands.

\begin{table*}[!ht]
    \caption{Summary statistics of hypergraphs generated by different real-world datasets. 
    The number of nodes $N$, 
    the number of hyperedges $M$, 
    the average degree $\left\langle k \right\rangle$,
    the average hyperdegree $\left\langle k^H \right\rangle$,
    the average cardinality of hyperedges $\left\langle k^E \right\rangle$ 
    and the value of $M/N$ of each hypergraph are given. In addition, we also show
    the average clustering coefficient $C$, 
    the average path length $\left\langle l \right\rangle$,
    and the density (refers to the link density) of the corresponding ordinary networks.}
    \resizebox{\textwidth}{!}{
    \centering
    \begin{tabular}{cccccccccc}
    \hline 
        {Network}       & $N$   & $M$   & $\langle k \rangle$ & $\left\langle k^H \right\rangle$  & $\left\langle k^E \right\rangle$ &  M/N   &     C & $\left\langle l \right\rangle$ & Density\\
    \hline
email-Enron     &   143 &  1459 &   36.26 &     31.94 &   3.13 &  10.2 &  0.66 &   1.9 &    0.25 \\
Algebra         &   423 &  1268 &   78.90 &     19.53 &   6.52 &  3.00 &  0.79 &  1.95 &    0.19 \\
Restaurants-Rev &   505 &   601 &    8.14 &      8.14 &   7.66 &  1.19 &  0.54 &  1.98 &    0.14 \\
Geometry        &   580 &  1193 &  164.79 &     21.52 &  10.47 &  2.06 &  0.82 &  1.75 &    0.28 \\
Music-Rev       &  1106 &   694 &  167.88 &      9.49 &  15.13 &  0.63 &  0.62 &  1.99 &    0.15 \\
NDC-classes     &  1161 &  1088 &   10.72 &      5.55 &   5.92 &  0.94 &  0.61 &   3.5 &    0.01 \\
Bar-Rev         &  1234 &  1194 &  174.30 &      9.62 &   9.93 &  0.97 &  0.58 &   2.1 &    0.14 \\
iAF1260b        &  1668 &  2351 &   13.26 &      5.46 &   3.87 &  1.41 &  0.55 &  2.67 &   0.01 \\
    \hline
    \end{tabular}
    }
    \label{tab:datasets}
\end{table*}

\section{Model Description}
\label{Model Description}
Gravity model, taking local and global topological characteristics of node into consideration, has been proved to be able to identify influential nodes both efficiently and accurately~\cite{zhao2021identification, li2019identifying, li2021identifying, bi2021temporal, cheng2015gene}.
Here, we propose a gravity-based centrality (HGC) method in hypergraphs, which incorporates the degree and the higher-order distance of nodes. 
Furthermore, a semi-local HGC (LHGC) is proposed to attain a trade-off between computational cost and accuracy.
In this section, we will first give basic definitions used in our methods, and then introduce the definitions of HGC and LHGC.

\subsection{Basic Definitions}
\label{basic definitions}
\renewcommand{\labelitemi}{$\bullet$}

\textbf{Distance between hyperedges.}
Two hyperedges are considered to be $s$-adjacent if they share at least $s$ nodes.
An $s$-walk with length $l$ is a sequence of successive $s$-adjacent hyperedges~\cite{aksoy2020hypernetwork}, defined by the following sequence:
\begin{equation}
\label{sWalk}
    \{(e_{n,0}, e_{n,1}),\ (e_{n,1},e_{n,2}),\cdots,(e_{n,l-1},e_{n,l})\},
\end{equation}
where $|e_{n,i-1}  \cap e_{n,i}| \ge s,\ i = 1,\cdots,l$, and $n$ is the sequence number of paths between $e_{n,0}$ and $e_{n,l}$.
The $s$-distance $d_{s}^e(g, q)$ between hyperedges $e_g$ and $e_q$ is the length of the shortest $s$-walk(s) between them.
Note that hyperedges are considered to be mutually unreachable if no such $s$-walk exists, and the $s$-distance between them is denoted as $\infty$. 

$1$-distance ($s=1$) is actually equivalent to the distance defined on ordinary networks, which is vastly used in the past works. 
However, the higher-order $s$-distance ($s \ge 2$) is so pervasive in real world that one cannot neglect it.
figure 2 shows the $s$-distance distribution of eight hypergraphs generated by real-world data, where $s$ ranges from $1$ to $9$.
The $s$-distance distributions are very similar across various $s$, and the proportion of hyperedge pairs decreased significantly with the increase of the value of the s-distance.
Actually, the maximum value of $s$-distance between hyperedge pairs (denoted as $s_m$ in the following context) varies in hypergraphs, which ranges from $9$ (iAF1260b) to $58$ (Geometry).
But we can still observe the prevalence of the higher-order $s$-distance in hypergraphs, even if $s$-distance where $s \ge 10$ is excluded here for simplicity.

\begin{figure*}[htp]
    \centering
    \includegraphics[width=\textwidth]{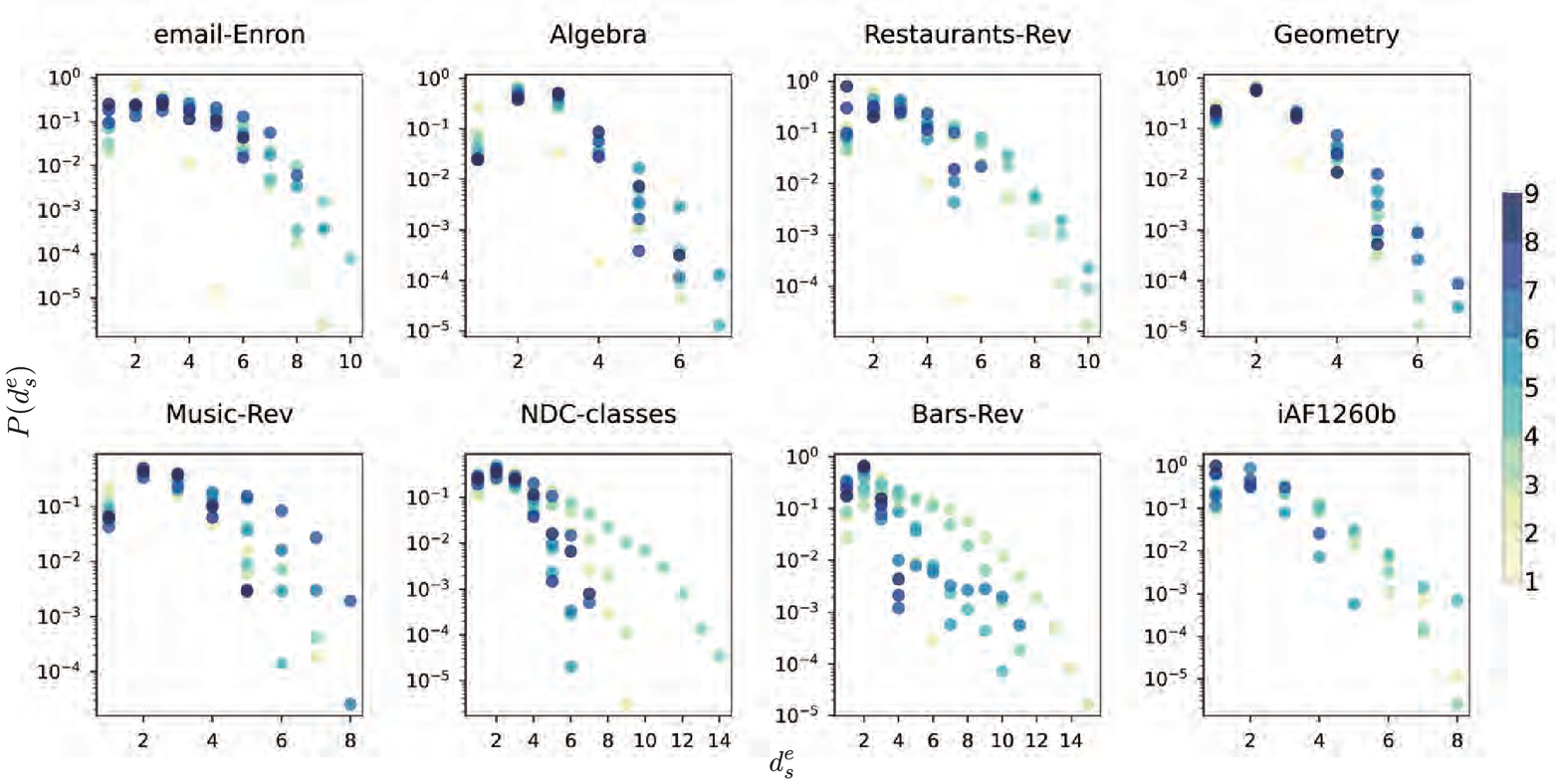}
    \caption{
    Distribution of $s$-distance, where $s \in [1, 9]$.
    The abscissa represents the value of $s$-distance $d_s^e$, and the ordinate represents the proportion of $d_s^e$.
    The colors of scatters from red to purple correspond to the values of $s$, as shown in the color bar.
    The inset clarifies the partial enlargement of each figure.
    } 
\label{fig:distance_distribution}
\end{figure*}

\textbf{Distance between nodes.}
Suppose nodes $v_i$ and $v_j$ belong to hyperedges $e_g$ and $e_q$ respectively, and we denote the $s$-distance between the two hyperedges as $d_{s}^e(g, q)$. 
The $s$-distance $d^v_{s}(i, j)$ between $v_i$ and $v_j$ is given as:
\begin{equation}
\label{NodeDistance}
    d_{s}^{v}(i, j)=
                    \begin{cases} 
                    1 & \text {, if } v_i,\ v_j \text{ exist in the same hyperedge} \\
                    d_{s}^{e}(g, q) + 1 & \text {, otherwise}
                    \end{cases}
\end{equation}
It should be noted that the $s$-distance between node $v_i$ and $v_j$ is set to be $N+1$ if the two nodes are not reachable through any $s$-walk. Thereafter, the higher-order distance $d^H(i,j)$ between $v_i$ and $v_j$ can be defined as follows:
\begin{equation}
\label{HigherOrderDistance}
    d^H(i,j)=\sum_{s=1}^{s_m} \alpha d_s^v(i, j),
\end{equation}
where $\alpha$ is a penalizing factor of the parameter $s$, here we set $\alpha=1/s^2$.
Consequently, $s$-distance with larger $s$ is considered to numerically contribute less to the definition of higher-order distance compared to the smaller $s$.

\subsection{Gravity-based Centrality in Hypergraphs (HGC)}
Given the above definition of higher-order distance, we further define the HGC score of a node $v_i$ as:

\begin{equation}
\label{HGM}
    G^H(i)=\sum_{i \ne j} \frac{k_i  k_j}{(d^H(i,j))^2},
\end{equation}
where $k_i$ is the degree of node $v_i$, and $d^H(i,j)$ is the higher-order distance between $v_i$ and $v_j$.
A node will be allocated a higher score if it has more neighbors locally and is more accessible to others globally. 

\subsection{Local Gravity-based Centrality in Hypergraphs (LHGC)}
HGC considers the local and global properties of nodes at the same time, and depicts the higher-order interactions by introducing higher-order distance.
But it is a costly job to calculate higher-order distance between nodes in a large-scale hypergraph.
Meanwhile, information from long path may bring noise to node ranking, since the influence between a node pair decays with the distance between them~\cite{li2019identifying, yang2020adaptive}. 
Consequently, we introduce LHGC, a semi-local version of HGC, for a node as follows:

\begin{equation}\label{LHGM}
    G^H_L(i)=\sum_{d^H(i,j) \le r_i, i \ne j} \frac{k_i  k_j}{(d^H(i,j))^2},
\end{equation}
where $r_i$ represented the radius of a node's valid influence.
Here, we set $r_i$ as the half of the largest higher-order distance from node $v_i$ to other nodes~\cite{Zhang2021LFIC}.
The computational complexity is reduced by introducing the cut-off parameter $r_i$, since the less influential paths are excluded from calculation.

\section{Benchmarks and Evaluation Metrics}
\label{evaluation}
In this section, we start by introducing the benchmark metrics that were proposed to characterize node importance in a hypergraph.
Then, we illustrate two evaluation methods, namely, SIR spreading influence and network $s$-efficiency, that will be used to evaluate the effectiveness of our centrality methods in various hypergraphs. 

\subsection{Benchmark Metrics}

\renewcommand{\labelitemi}{$\bullet$}

 \textbf{DC:}
Degree centrality is equivalent to the degree of the ordinary network, which takes the number of neighbors of a node as its centrality score. 
The degree of a node $v_i$ can be calculated by:
\begin{equation}
    k_i=\sum_{j=1}^N A_{ij},
\end{equation}
where $A$ is the adjacency matrix of the hypergraph.

 \textbf{HDC:}
Hyperdegree centrality assumes that the more hyperedges a node is incident with, the more vital the node is.
The hyperdegree of a node is formally defined as:
\begin{equation}\label{HDC}
    k^H_i=\sum_{m=1}^M B_{im},
\end{equation}
where $B$ is the incidence matrix of the hypergraph.

 \textbf{VC:}
Vector centrality evaluates the importance of a node in terms of a vector, where the $k_{th}$ component describes the importance of the node in a hyperedge with size $k+1$. 
Here, $k = 1,\ \cdots ,\ K-1$, where $K=\max_{m=1}^M\{ |e_m|\}$ corresponds to the maximum cardinality of hyperedges~\cite{kovalenko2021vector}.
It first projects a hypergraph into line graph~\footnote{
In particular, the line graph $L(H)$ is a graph of $M$ nodes. A node and a hyperedge in $H=(V, E)$ are mapped to an edge and a node in $L(H)$, respectively.
That is to say, there is an edge between node $v_g$ and $v_q$ in $L(H)$ if and only if $e_g \cap e_q \ne \emptyset$ in $H$.}.
Then, it calculates the eigenvector centralities of all hyperedges, which is denoted as $c(m),\ m=1,\ \cdots,\ M$. Finally, the vector centrality $\overrightarrow{c_{i}}$ of node $v_i$ is obtained by:
\begin{equation}
    \overrightarrow{c_{i}}=\left(c_{i 2}, \cdots, c_{i K}\right)^{\mathrm{T}} \in \mathbb{R}^{K-1},
\end{equation}
where $c_{i k}=\frac{1}{k} \sum_{\substack{i \in e_m \\ |e_m|=k}} c(m)$, indicating the score of a hyperedge is distributed evenly to its nodes. 
VC evaluates node importance by a vector.
In this paper, we take the sum of elements in $\overrightarrow{c_i}$ as the final VC centrality score of node $v_i$:
\begin{equation}
    V(i)=\sum_{k=2}^K c_{ik}
\end{equation}

In this following, we define three kinds of node centrality, i.e., HEDC, ECC and HCC, based on the centrality of a hyperedge. To achieve this, we first define the centrality of a hyperedge, and then the centrality score of each hyperedge is evenly distributed to the nodes that belong to the same hyperedge.

 \textbf{HEDC:}
In the line graph of a hypergraph, hyperedges are adjacent if they share at least one node. Hence, the degree of a hyperedge can be defined as the number of hyperedges that it is adjacent to.
Hyperedge degree centrality assumes that a hyperedge is important if it has large degree~\cite{hu2021aging, wang2010evolving}.
We denote the adjacency matrix of a line graph as $A^L$.
$A^L_{f g}=1$, if hyperedges $e_f$ and $e_g$ are adjacent, otherwise $A^L_{f g}=0$.
$A_{ff}$ is set to 0.
The hyperedge degree centrality (HEDC) of a hyperedge $e_m$ (denoted as $S^e_{HEDC}(m),\ m=1,\cdots,M$)  is defined as:
\begin{equation}
    S^e_{HEDC}(m) = \sum_{g=1}^M A^L_{mg}
\end{equation}
Therefore, we can obtain the HEDC centrality of a node $v_i$ by distributing the centrality score of hyperedge $S^e_{HEDC}(m)$ evenly to its nodes:

\begin{equation}
    S^v_{HEDC} (i)=\sum_{m=1}^M B_{im}  \frac{S^e_{HEDC}(m)}{|e_m|} 
\end{equation}

 \textbf{ECC:}
$s$-eccentricity considers that the shorter $s$-distance of a hyperedge to other hyperedges in the network, the more important the hyperedge is~\cite{aksoy2020hypernetwork}.
The $s$-eccentricity of a hyperedge is the maximum $s$-distance to other hyperedges in the $s$-connected component, which is defined as:
\begin{equation}
    S_{ECC}^e(g)=\max_{e_q \in C_S} \{d^e _s(g, q)\},
\end{equation}
where $C_s$ denotes the $s$-connected component that consists of $s$-connected hyperedges.
We calculate $S_{ECC}^e(g)$ by using $s=1$ in our experiments.
Similar to HEDC, the $s$-eccentricity of a node $v_i$ (ECC) is obtained by the following equation:

\begin{equation}
    S^v_{ECC} (i)=\sum_{m=1}^M B_{im}  \frac{S^e_{ECC}(m)}{|e_m|} 
\end{equation}

 \textbf{HCC:}
The $s$-harmonic closeness centrality posits that the smaller the average $s$-distance of a hyperedge to other hyperedges, the more important the hyperedge is~\cite{aksoy2020hypernetwork}.
The $s$-harmonic closeness centrality score of a hyperedge is defined as:
\begin{equation}
    S_{HCC}^e(g) = \frac{1}{|E_s|-1}\sum\limits_{\substack{e_g, e_q \in E_s \\ g \ne q}}\frac{1}{d^e_s(g, q)},
\end{equation}
where $E_s=\{ e_m: |e_m| \ge s \}$ is the set of hyperedges that contains at least $s$ nodes.
We calculate $S_{HCC}^e(g)$ by setting $s=1$ in our experiments.
Similarly, the $s$-harmonic closeness centrality (HCC) of a node $v_i$ is given by:

\begin{equation}
    S^v_{HCC} (i)=\sum_{m=1}^M B_{im} \frac{S^e_{HCC}(m)}{|e_m|} 
\end{equation}

\subsection{SIR Spreading Dynamics in Hypergraphs}
The contagion through pairwise interactions is prevalent in real world, such as the disease spreading and information diffusion under domino effects, which is often referred to as simple contagion.
However, the pairwise interactions are not enough to characterize social contagion processes, where more complicated mechanisms of influence and reinforcement are at work, such as peer pressure and reinforcement of atmosphere~\cite{iacopini2019simplicial, jhun2019simplicial}. We refer to this kind of contagion as complex contagion process~\cite{jhun2019simplicial, landry2020effect, de2020social, jhun2021effective}.
In fact, understanding groups is a critical aspect to gain insight on individual behaviors.
Generally, most human behavior is easily influenced by their groups' behavior, and most of the world's decisions or work are the results of groups or teams.
Taking information diffusion as an example, it usually starts with a particular spreader and several related social groups. The members in one group will spread information to other groups if the information has gained a fraction of attention in the group. 
Here, we introduce a parameter $\eta$ (a threshold value) to represent how much attention the information has gained that makes the members spread it to other groups.

We use susceptible-infected-recovered (SIR) model with the threshold value $\eta$ to mimic a complex contagion process in a hypergraph to assess the performance of the centrality methods proposed above.
In the SIR model, a node can be in one of the three states, i.e., susceptible ($S$), infected ($I$) or recovered ($R$).
The spreading process is described as follows: 
\begin{enumerate}
\item \textbf{Initialization.}
A node $v_i$ is randomly chosen as the initial seed of the spreading process, i.e., in the state of I.
\item \textbf{Contagion.}
At the first time step, the seed node $v_i$ will infect the S nodes in the hyperedges in which $v_i$ is located with infection probability $\beta$. For each of the hyperedges that contain $v_i$, e.g., $e_m$, if the fraction of infected nodes (i.e., I and R nodes) is equal to or larger than $\eta$, the infected nodes in $e_m$ will infect the S nodes in the hyperedges that are adjacent to $e_m$ at the coming step.
Such infection and expansion process will be repeated at each time step. 
\item \textbf{Recovery.}
At each time step, every I-state node in the hypergraph recovers with probability $\gamma$ independently.
\item \textbf{Termination.} The contagion and recovery processes will stop after $T$ steps, where $T$ is a control parameter.
\end{enumerate}

\begin{figure}
    \centering
    \includegraphics[width=\textwidth]{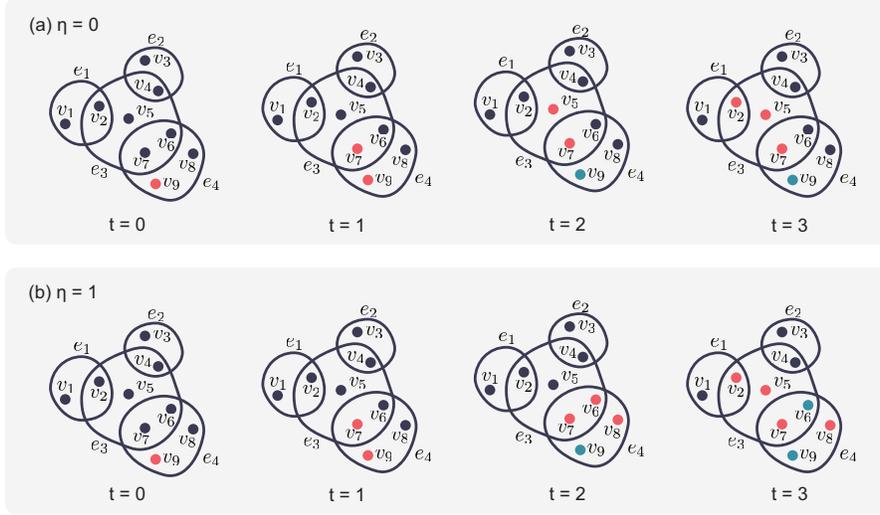}
    \caption{An example of spreading processes within $4$ time steps with (a) $\eta=0$ and (b) $\eta=1$.
    The black, red and green nodes correspond to S, I, R nodes, respectively.}
    \label{fig:spread_exm}
\end{figure}

To concretely illustrate the complex contagion process we proposed above, we give an example in figure~\ref{fig:spread_exm}, where $\eta=0$ and $1$, respectively. 
We use black, red and green color to present susceptible, infected and recovered nodes, respectively, and we use the same seed node $v_9$ for different $\eta$. 
By setting $\eta=0$ (as shown in figure~\ref{fig:spread_exm}(a)), we don't consider the group effect, that is to say, the spreading is the same as that on an ordinary network. Specifically, $v_9$ is selected as the seed and activated at time step $t=0$.
When $t=1$, $v_9$ infects one S-state neighbor ($v_7$), and hyperedges $e_3,\ e_4$ reaches the spreading threshold $\eta=0$.
When $t=2$, the node $v_5$ is infected, and $v_9$ is recovered.
When $t=3$, $v_2$ is infected and turns into I state.
When $\eta=1$ (as show in figure~\ref{fig:spread_exm}(b)), the suppression from a group (or a hyperedge) is maximized, and an I-state node has a chance to infect susceptible neighbors from other groups (or hyperedges) only when all nodes in the current group (or hyperedge) are infected.
The fraction of I and R nodes of the hyperedges $e_4$ do not reach the threshold of $1$ until $t=2$.
The infected nodes in $e_4$ start to infect S nodes in its incident hyperedge (that is, the hyperedge $e_3$) at time step $t=3$.

To obtain the spreading influence of a node $v_i$ at time step $T$, we conduct the above SIR model in a hypergraph by setting $v_i$ as the seed node. The expected spreading influence of a node $v_i$ is the average over 50 times of Monte Carlo simulations.

\subsection{Network $s$-Efficiency}
\label{sec:4.3}
Spreading model depicts how vital a node is from the perspective of dynamics, while the network efficiency can reflect the importance of a node functionally. 
The network efficiency is an indicator that quantifies the efficiency of information exchange in a network.
It assumes that the more distant two nodes are, the less efficient their communication would be.
And the network efficiency of a hypergraph can be described as the $s$-efficiency $\mathcal{E}_s$~\cite{aksoy2020hypernetwork}, which can be calculated as:

\begin{equation}
\label{Efficiency}
    \mathcal{E}_s = \left(\begin{array}{c}
\left|E_{s}\right| \\
2
\end{array}\right)^{-1} \sum_{\substack{e_g, e_q \in E_{s} \\ g \neq q}} \frac{1}{d_{s}^{e}(g, q)},
\end{equation}
where $E_s=\{ e_m: |e_m| \ge s \}$ is the set of hyperedges that contains at least $s$ nodes.
So the efficiency of a hypergraph will decrease dramatically if we remove vital nodes from a hypergraph.
To start with, we calculate the $s$-efficiency of a hypergraph (denoted as $\mathcal{E}_s$).
Then, we attack $p$ proportion of top nodes ranked by a particular centrality measure, and recalculate the $s$-efficiency of the remaining hypergraph, which is denoted as $\mathcal{E}_s(p)$.
The efficiency loss between $\mathcal{E}_s$ and $\mathcal{E}_s(p)$ is denoted as $\Delta \mathcal{E}_s(p)$, which is used to quantify the importance of the deleted nodes.
This process allows us to compare the ability of centrality measures in terms of identifying nodes that are critical to hypergraph connectivity.
It is noteworthy that 1-efficiency is just the equivalence of network efficiency defined on ordinary networks, while higher-order efficiency is more resilient than the pairwise ones, thus is more valuable to maintain high network efficiency. 
As a result, we consider $\mathcal{E}_s(p)$ with different $s$ to quantify the efficiency loss.
The sum of $\Delta \mathcal{E}_s(p)$ with different $s$ ($\Delta \mathcal{E}(p)$) is defined as:
\begin{equation}
    \Delta \mathcal{E}(p) = \sum_{s=1} ^ {s^{\prime}} \Delta \mathcal{E}_s(p)
\end{equation}
Here, $s^{\prime}$ is a tunable parameter.
In this paper, we conduct the experiments by taking $s^{\prime}$ as $3,\ 6,\ 9$, since the maximum values of $s^{\prime}$ of each hypergraph can take ($s_m$) are different.
$\mathcal{E}_s(p)$ will drop significantly if the nodes we delete are vital.
Thus, larger efficiency loss (denoted as $\Delta \mathcal{E}(p)$) indicates a better performance.

\section{Results}
\label{Results}
\subsection{Node spreading influence quantification}
\label{sec:5.1}
\begin{figure*}[!htb]
    \centering
    \includegraphics[width=\textwidth]{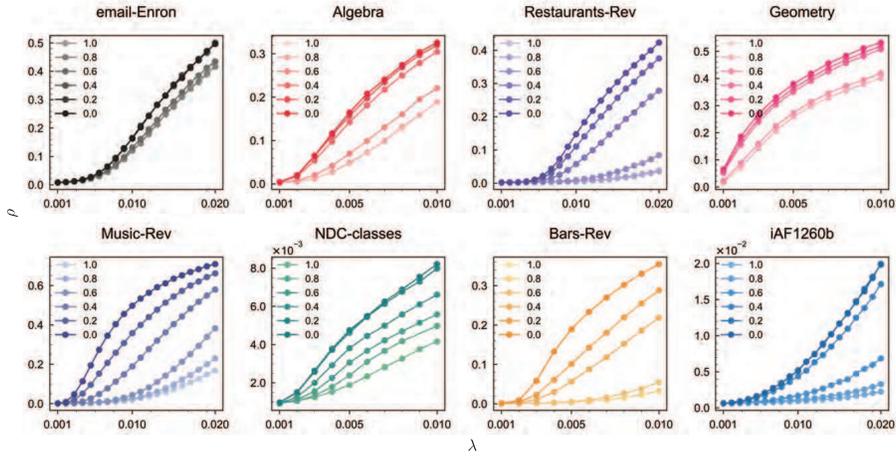}
    \caption{
    The fraction of infected $\rho$ under different $\lambda$, where $\lambda=\frac{\beta}{\gamma}$. We set $\gamma=1$.
    The infection probability $\beta$ and the threshold value $\eta$ are tuned. Each of the points in the figures are averaged over $50$ times of simulations.
    }
\label{fig:params_tuning}
\end{figure*}

To quantify the node spreading influence, we first conduct numerical simulation for SIR spreading dynamics in hypergraphs generated by real-world data. We denote the effective spreading rate as $\lambda=\frac{\beta}{\gamma}$. 
In the simulation we set $\gamma=1$, and tune the infection probability $\beta$ with $\eta$ ranging from $0$ to $1$ with an interval of $0.2$. In figure~\ref{fig:params_tuning}, we observe that most hypergraphs reach a considerable fraction of infected (denoted as  $\rho$, which is computed by the fraction of infected and recovered at time step $T=10$) with $\lambda$ no less than 0.01, while the fraction of infected of NDC-classes and iAF1260b are relatively small compared to the others.
Unsurprisingly, $\rho$ grows slower and reaches a smaller scale with the increase of $\eta$ due to the suppression effect introduced by $\eta$. For each of the hypergraphs, we choose $\lambda$ slightly larger than the spreading threshold $\lambda_c$ when quantifying the spreading influence of each node. 
The values of $\lambda$ that are used in our experiments under different $\eta$ are illustrated in Table~\ref{tab:params_tuning}.

\begin{table*}[htb]
    \centering
    \caption{The values of $\lambda$ that are used in our experiments under different $\eta$.}
    
    {
    \centering
    \begin{tabular}{c|cccccc}
    \hline
    \diagbox{Network}{$\eta$} &    0.0 &    0.2 &    0.4 &    0.6 &    0.8 &   1.0 \\
    \hline
    email-Enron       &  0.005 &  0.010 &  0.010 &  0.011 &  0.011 &  0.011 \\
    Algebra           &  0.003 &  0.004 &  0.004 &  0.006 &  0.007 &  0.007 \\
    Restaurants-Rev   &  0.009 &  0.011 &  0.014 &  0.020 &  0.030 &  0.030 \\
    Geometry          &  0.001 &  0.001 &  0.001 &  0.002 &  0.002 &  0.003 \\
    Music-Rev         &  0.003 &  0.005 &  0.009 &  0.014 &  0.015 &  0.016 \\
    NDC-classes       &  0.003 &  0.003 &  0.003 &  0.004 &  0.005 &  0.006 \\
    Bars-Rev          &  0.004 &  0.005 &  0.006 &  0.010 &  0.011 &  0.012 \\
    iAF1260b          &  0.005 &  0.005 &  0.005 &  0.010 &  0.015 &  0.025 \\
    \hline
    \end{tabular}
    }
    \label{tab:params_tuning}
\end{table*}

We test the performance of our centrality methods in  identifying early-time influencers for SIR spreading dynamics in hypergraphs.
Generally speaking, decision makers care more about the top-ranked nodes. 
In view of this demand, we quantify the spreading influence of the top $10\%$ ranked nodes by each centrality metric, and plot the average spreading curves within $T=5$ steps for different hypergraphs and different values of $\eta$. 
We find that each centrality method performs relatively consistent with the change of $\eta$. 
Overall speaking, the gravity-based methods, i.e., HGC and LHGC, perform better when $\eta$ is larger, reflecting the consistency of our analysis and the superiority of our methods with respect to non-pairwise spreading process. 
We take $\eta=0.2$ as the example to illustrate the results, and the results for other $\eta$ can be found in figures S1-S5 in the Appendix. figure~\ref{fig:speed_0} shows the spreading curves of the top $10\%$ ranked nodes by each centrality metric when $\eta=0.2$. 
HGC and LHGC outperform the baselines (or perform similarly to the second best baseline) in most hypergraphs, namely Restaurants-Rev, Music-Rev, Bars-Rev and iAF1260b (Algebra and Geometry). 
The curves of HGC are invisible because its results are close to  LHGC. 

The detailed observations of the baselines are as follows. 
Firstly, we notice that the spreading curves of HGC and LHGC is close to DC in most hypergraphs, and is even better than DC in NDC-classes.
This is because DC is highly correlated with HGC, with Pearson Correlation Coefficient (PCC) higher than 0.9 in all the hypergraphs, as shown in  figure~\ref{fig:mc_email}(a) for email-Enron (the corresponding results for other data are shown in figures S6-12 in the Appendix).
It should be noted that ECC and HCC are methods based on the paths defined on ordinary networks, i.e., they only consider 1-walk in a hypergraph. They perform worse than the other methods, indicating that we need to consider the higher-order distance between nodes for vital node identification in a hypergraph. HDC, which allocates nodes with larger hyperdegree higher centrality scores, performs relatively mediocre. It suggests that nodes exist in the intersection of hyperedges  may be less important for influence spreading. In the definition of hyperedge based methods (including HEDC, ECC and HCC) and VC, the hyperedge centrality score is evenly distributed to each of the nodes in the hyperedge. The bad performance of those methods implies that nodes may contribute unevenly to their incident hyperedges. 

\begin{figure*}[htp]
    \centering
    \includegraphics[width=\textwidth]{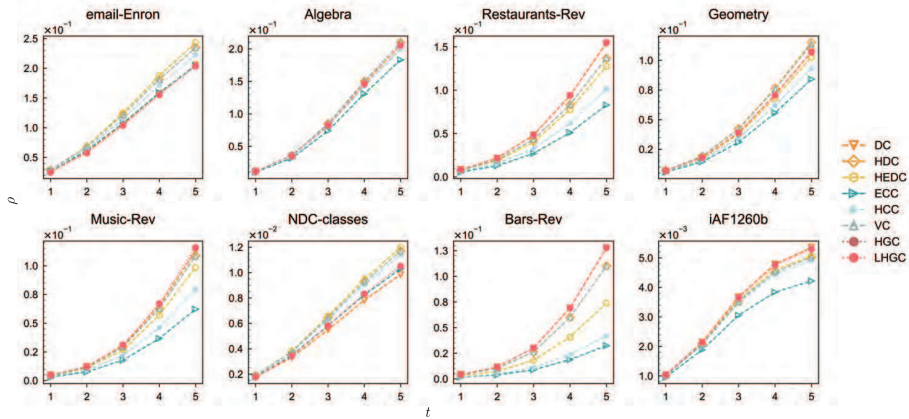}
    \caption{
   Spreading curves of the top $10\%$ ranked nodes by each centrality metric when $\eta=0.2$. 
    }
\label{fig:speed_0}
\end{figure*}

To analyze the relationship between the centrality metrics and their ability of mining influential nodes in more detail, we show the correlation between HGC and other baselines metrics in  figure~\ref{fig:mc_email} as well as figures S6-S12 in the Appendix for all the hypergraphs. The color in each of the figures reveals the spreading ability of the nodes, with color changing from cool to warm representing nodes have low to high spreading ability. The spreading ability of each node is quantified by the expected influence at time step $t=5$ by setting each node as the seed. The spreading parameters are the same as those of figure~\ref{fig:speed_0}. It should be noted that we only show the results of $\eta=0.2$, as the results are consistent while using different $\eta$. We observe that HGC is highly correlated with DC in all the hypergraphs, with PCC higher than $0.9$, but the correlation between HGC and other metrics is relatively low. In addition, the results show that nodes that are scored a high value by HGC are generally influential nodes indeed (red or yellow nodes), while other methods show instabilities. The details are as follows. In email-Enron (figure~\ref{fig:mc_email}), even though HDC, HEDC and VC perform better than our methods in general (as shown in figure~\ref{fig:speed_0}), the low-influential nodes, such as those colored in purple, are given relatively dispersed scores by these methods. The similar patterns can also be found in other methods, namely ECC and HCC, as well as hypergraphs, e.g., Restaurants-Rev, Music-Rev and Bars-Rev etc.
(figures S6-S12 in the Appendix). 

\begin{figure*}[!htb]
    \centering
    \includegraphics[width=\textwidth]{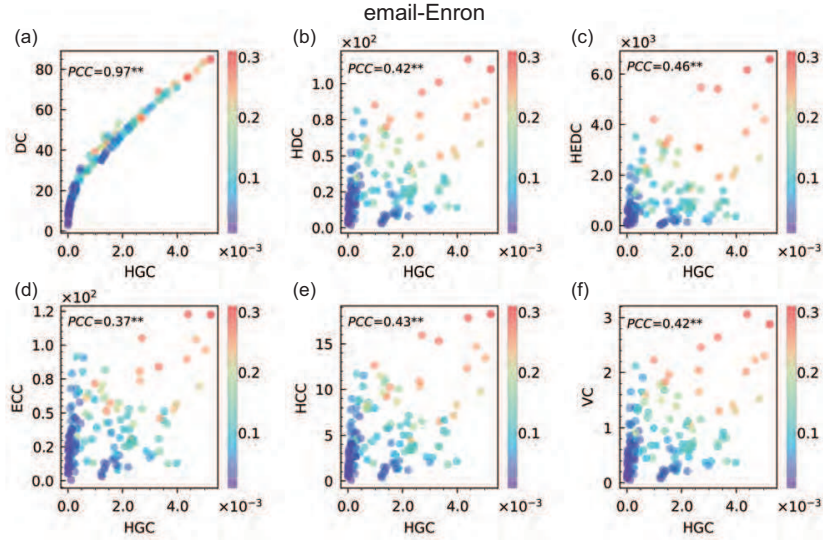}
    \caption{
    The correlation between HGC and different benchmarks in hypergraph email-Enron. The color of point represents the spreading ability of each node, with color changing from cool to warm representing nodes have low to high spreading ability. The $x$ axis and $y$ axis are the node scores obtained by different centrality metrics.
    PCC is the Pearson Correlation Coefficient between two metrics, ** denote the p-value is smaller than 0.05.
    }
\label{fig:mc_email}
\end{figure*}

\begin{figure*}
    \centering
    \includegraphics[width=0.5\textwidth]{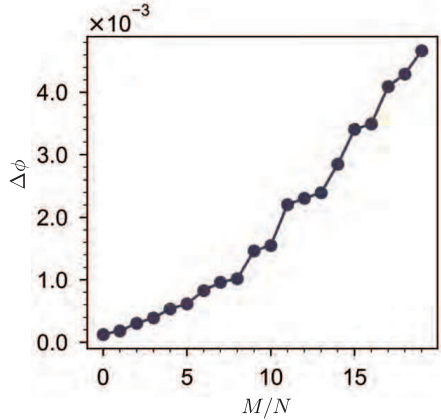}
    \caption{The change of $\Delta \phi$ when the value of $M/N$ is altered. The synthetic hypergraphs are generated by HyperCL.
    For each value of $M/N$, the hypergraphs are generated independently 50 times, and the SIR spreading dynamics on each hypergraph is conducted another 50 times independently.}
    \label{fig:artificial_network}
\end{figure*}

We further inspect the topology of our datasets in retrospect, and notice that the hypergraphs on which our methods show good performance tend to have a low ratio of $M/N$ (as shown in Table~\ref{tab:datasets}).
To further verify our observation, we alter $M/N$ in different synthetic hypergraphs generated by HyperCL~\cite{lee2021hyperedges}, which is a random hypergraph generator designed to generate hypergraphs with a certain hyperdegree distribution. In the generation of synthetic hypergraphs, we keep the hyperdegree distribution unchanged and set the number of nodes as $N=1000$. 
The number of hyperedges $M$ is tuned. For each hypergraph generated by a specific value of $M$, we first conduct the SIR spreading dynamics by setting each node as the seed and find the top $10\%$ of nodes that have the largest spreading influence within 5 time steps ($\beta=0.01$, $\gamma=1$, $\eta=0.2$). 
The average area under the spreading curve of these top $10\%$ nodes is denoted as $\phi_0$, indicating the average spreading capacity of them. 
We denote the average area under the spreading curve of top $10\%$ nodes that are ranked by HGC as $\phi$. 
figure~\ref{fig:artificial_network} shows the change of $\Delta \phi=\phi_0-\phi$ under different values of $M/N$.
As $M/N$ decreases, the spreading capacity difference $\Delta \phi$ of HGC decreases, which further suggests that HGC can better identify high influential nodes in hypergraphs with lower value of $M/N$.
When there are fewer nodes than hyperedges, the hypergraph is generally dense, which means nodes can reach each other within a few hops, and the contribution of structural information of paths will be less.

\clearpage
\subsection{Network Efficiency quantification}
We evaluate the performance of our centrality methods in finding nodes that are important in terms of hypergraph connectivity in this part. We first delete a fraction of $p$ top ranked nodes by each of the centrality metric, and then compute the efficiency loss $\Delta \mathcal{E}(p)$ (defined in Section~\ref{sec:4.3}). 

\begin{figure}[!htb]
    \centering
    \includegraphics[width=\textwidth]{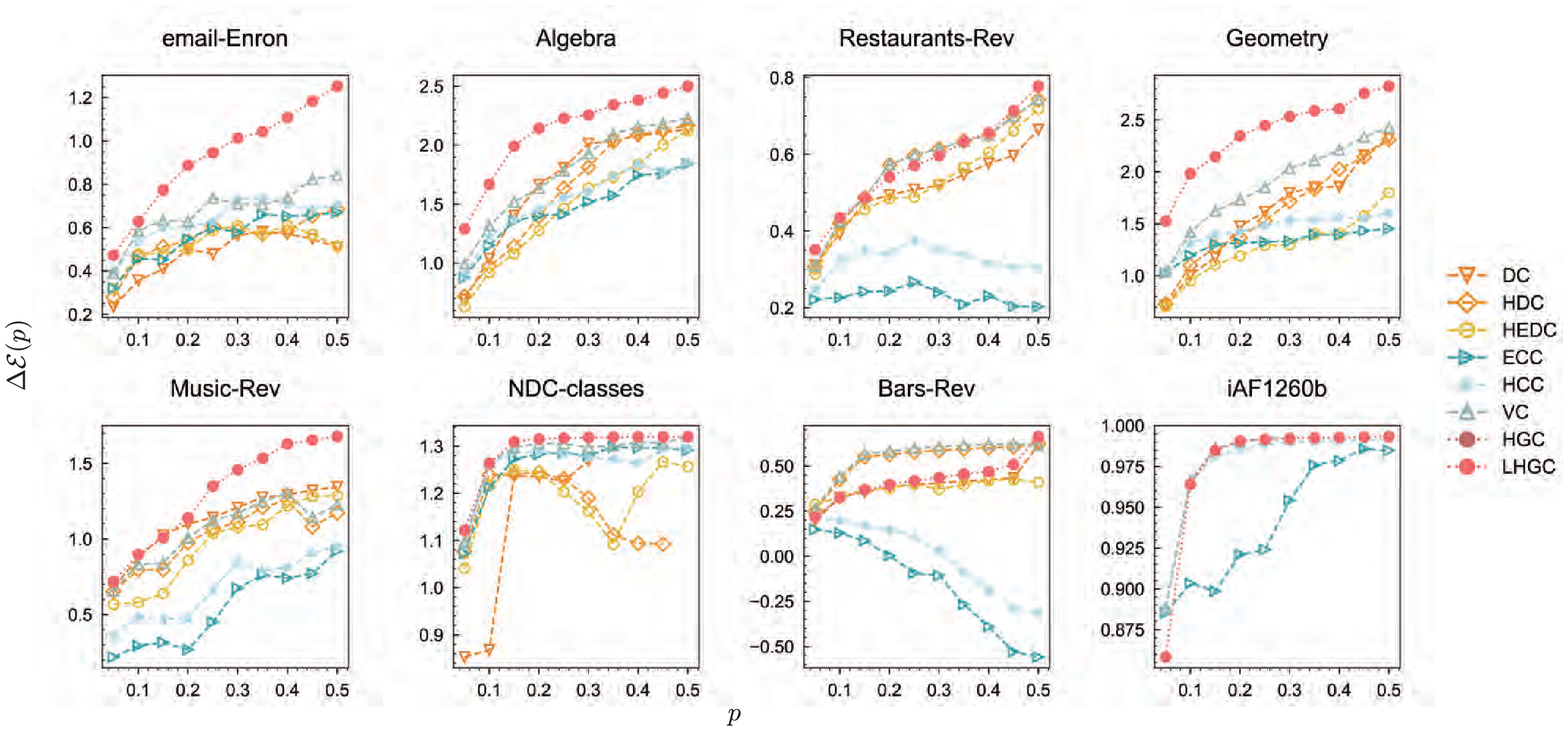}
    \caption{
    The change of $\Delta \mathcal{E}(p)$ by attacking nodes.
    The $x$ axis denotes the proportion of top ranked nodes we delete from a hypergraph, and the $y$ axis denotes the $\Delta \mathcal{E}(p)$, where $s^{\prime}=6$.
    }
\label{fig:efficiency}
\end{figure}

We take figure~\ref{fig:efficiency} as an example ($s^{\prime}=6$). Results for $s^{\prime}=3$ and $9$ are given in figures S13-14 in the Appendix, which are consistent with $s^{\prime}=6$. figure~\ref{fig:efficiency} shows the change of $\Delta \mathcal{E}(p)$ after the attack on $p$ $(p \in [0.05, 0.5])$ fraction of top ranked nodes.
Overall, we observe that our methods, i.e., HGC and LHGC, are superior to other centrality metrics significantly in most of the hypergraphs, i.e., email-Enron, Algebra, Geometry, Music-Rev, NDC-classes, and are only inferior in Bars-Rev. These observations suggest that HGC and LHGC can disentangle hypergraphs better by destructing higher order structures, that is to say, they are able to find nodes that play vital roles in improving network efficiency.
The remaining methods, on the other hand, are quite fluctuating. 
In contrast to the performance of node influence, DC is not as good as other metrics in terms of attacking nodes, indicating that DC may be able to screen out fast spreaders, but it fails to find nodes that play an important role in hypergraph connectivity.
The $\Delta \mathcal{E}(p)$ of ECC and HCC experience ups and downs in many hypergraphs, such as Algebra, Restaurants-Rev, Music-Rev, Bars-Rev and iAF1260b.
Counterintuitively, these two path-based metrics do not perform very well, and $\Delta \mathcal{E}(p)$ even becomes smaller when we remove nodes that are ranked as vital by them.
This shows that the nodes that they allocate a high centrality score may be redundant to improve the network efficiency, and this is not decision makers expect.
The values of $\Delta \mathcal{E}(p)$ of VC and HEDC grow fastest in Restaurants-Rev and Bars-Rev when $p$ is small, but our methods catches up when $p$ gets larger.

\clearpage
\section{Conclusions \& Discussion}
\label{Conclusions}
In this work, we proposed two new  centrality methods in hypergraphs based on the gravity model, namely, HGC and LHGC.
HGC incorporates both local topological characteristic (degree) and global path-based information (higher-order distance).
Furthermore, LHGC was proposed to achieve lower computational complexity while preserving the accuracy.

In this work, we evaluate the performance of different centrality metrics defined in hypergraphs comprehensively, while others are constrained within the evaluation metrics defined on ordinary network~\cite{aktas2021identifying}. We first proposed a complex contagion model, i.e., SIR model with a threshold value, to indicate the group effect in the hypergraphs. Then, we evaluate the performance of our methods in finding influential nodes based on the SIR model with threshold.
The analysis of the spreading dynamics shows that the proposed methods behave comparatively optimal or close to the optimal in most hypergraphs.
The correlation analysis between different centrality metrics microscopically unveils that our methods are able to distinguish the spreading influence of nodes more accurately compared with 6 other baselines.
In the sequel, we investigated the performance of our methods in finding influential nodes on synthetic hypergraphs generated by a hypergraph generator model (HyperCL), which shows that our methods give better performance in hypergraphs with a lower value of $M/N$, where $M$ and $N$ are the number of hyperedges and nodes, respectively.
Finally, the stability of the proposed methods in the qualification of the network efficiency suggests our methods are able to find out nodes that are vital to sustain hypergraph connectivity.

The two metrics we proposed are based on the higher-order distance between nodes, which have considered the higher-order information in a system. We deem that our work can boost the research in defining metrics for vital node identification in a hypergraph as we have defined two systematic evaluation metrics to evaluate the performance of different centrality metrics. Nevertheless, we should notice that there are still some potential in the future work.
First of all, entropy has been proved to be effective in quantifying node importance~\cite{nikolaev2015efficient, qiao2017identify}, thus combining the s-walk defined in hypergraph with entropy could be a promising way to find important nodes.
What's more, one may consider using semi-local information from neighbors, rather than just degree, e.g., the degree of the second-order neighbors, to define centrality methods. Last but not the least, the methods we proposed in this work may also shed light on some related problems, i.e., influence maximization~\cite{kundu2011new} and network dismantling~\cite{wang2020neighborhood, ren2019generalized}.


% ...

\section{Acknowledgments}
This work was supported by Natural Science Foundation of Zhejiang Province (Grant Nos. LQ22F030008 and LR18A050001), the National Natural Science Foundation of China (Grant Nos. 92146001, 61873080 and 61673151), the Major Project of The National Social Science Fund of China (Grant No. 19ZDA324), and the Scientific Research Foundation for Scholars of HZNU (2021QDL030).

{\bibliography{TempExample}}

\end{document}